\documentclass[conference]{IEEEtran}
\IEEEoverridecommandlockouts
\usepackage{amsmath,amsfonts}
\usepackage{algorithm}
\usepackage[noend]{algpseudocode}
\usepackage{textcomp}
\usepackage{soul}
\usepackage{booktabs}
\usepackage[utf8]{inputenc}
\usepackage[T1]{fontenc}
\usepackage{microtype}
\usepackage{rotating}
\usepackage{graphicx}
\usepackage{paralist}
\usepackage{tabularx}
\usepackage{multicol}
\usepackage{multirow}
\usepackage{pbox}
\usepackage{enumitem}	
\usepackage{colortbl}
\usepackage{pifont}
\usepackage{xspace}
\usepackage{url}
\usepackage{tikz}
\usepackage{tabularx}
\usepackage{fontawesome}
\usepackage{lscape}
\usepackage{listings}
\usepackage{color}
\usepackage{anyfontsize}
\usepackage{comment}
\usepackage{soul}
\usepackage{multibib}
\usepackage{tcolorbox}
\usepackage{adjustbox}
\usepackage{marginnote}
\usepackage{setspace}
\usepackage[colorlinks=true, allcolors=blue]{hyperref}

\newcommand{\ie}{\emph{i.e.,}\xspace}
\newcommand{\eg}{\emph{e.g.,}\xspace}
\newcommand{\etc}{etc.\xspace}
\newcommand{\etal}{\emph{et~al.}\xspace}
\newcommand{\secref}[1]{Section~\ref{#1}\xspace}
\newcommand{\figref}[1]{Fig.~\ref{#1}\xspace}

\newcommand{\tabref}[1]{Table~\ref{#1}\xspace}

\newcommand{\FC}{\emph{Full context}\xspace}
\newcommand{\NFC}{\emph{Non-full context}\xspace}
\newcommand{\copilot}{\emph{Copilot}\xspace}
\newcommand{\Gcopilot}{\emph{GitHub Copilot}\xspace}
\newcommand{\methods}{892\xspace}

\newcommand{\smalltexttt}[1]{{\small \texttt{#1}}}

\definecolor{gray50}{gray}{.5}
\definecolor{gray40}{gray}{.6}
\definecolor{gray30}{gray}{.7}
\definecolor{gray20}{gray}{.8}
\definecolor{gray10}{gray}{.9}
\definecolor{gray05}{gray}{.95}

\newlength\Linewidth
\def\findlength{\setlength\Linewidth\linewidth
	\addtolength\Linewidth{-4\fboxrule}
	\addtolength\Linewidth{-3\fboxsep}
}

\newboolean{showcomments}

\setboolean{showcomments}{true}

\ifthenelse{\boolean{showcomments}}{
  \marginparwidth=1cm
  \newcommand{\nb}[2]{
	\marginnote{\raggedright\tiny\textcolor{red}{{*\textit{#2} (#1)}}}\textcolor{red}{*}
  }
}
{
  \newcommand{\nb}[2]{}
}

\def\BibTeX{{\rm B\kern-.05em{\sc i\kern-.025em b}\kern-.08em
    T\kern-.1667em\lower.7ex\hbox{E}\kern-.125emX}}
    
\definecolor{lightergray}{rgb}{0.9,0.9,0.9}
\newtcolorbox{resultbox}{colback=lightergray, arc=0.5mm, top=2mm, bottom=2mm, left=2mm, right=2mm}

\begin{document}

\title{On the Robustness of Code Generation Techniques:\\An Empirical Study on GitHub Copilot}

\author{
\IEEEauthorblockN{Antonio Mastropaolo\IEEEauthorrefmark{1}, Luca Pascarella\IEEEauthorrefmark{1}, Emanuela Guglielmi\IEEEauthorrefmark{2}, Matteo Ciniselli\IEEEauthorrefmark{1}\\Simone Scalabrino\IEEEauthorrefmark{2}, Rocco Oliveto\IEEEauthorrefmark{2}, Gabriele Bavota\IEEEauthorrefmark{1}}
\IEEEauthorblockA{\IEEEauthorrefmark{1}\textit{SEART @ Software Institute, Universit\`{a} della Svizzera italiana (USI), Switzerland}}
\IEEEauthorblockA{\IEEEauthorrefmark{2}\textit{University of Molise, Italy}}
}

\maketitle

\begin{abstract}
Software engineering research has always being concerned with the improvement of code completion approaches, which suggest the next tokens a developer will likely type while coding. 
The release of GitHub Copilot constitutes a big step forward, also because of its unprecedented ability to automatically generate even entire functions from their natural language description. 
While the usefulness of Copilot is evident, it is still unclear to what extent it is robust. Specifically, we do not know the extent to which semantic-preserving changes in the natural language description provided to the model have an effect on the generated code function.
In this paper we present an empirical study in which we aim at understanding whether \emph{different but semantically equivalent natural language descriptions result in the same recommended function}. A negative answer would pose questions on the robustness of deep learning (DL)-based code generators since it would imply that developers using different wordings to describe the same code would obtain different recommendations.
We asked Copilot to automatically generate \methods Java methods starting from their original Javadoc description. Then, we generated different semantically equivalent descriptions for each method both manually and automatically, and we analyzed the extent to which predictions generated by Copilot changed. Our results show that modifying the description results in different code recommendations in $\sim$46\% of cases. Also, differences in the semantically equivalent descriptions might impact the correctness of the generated code ($\pm$28\%).
\end{abstract}


\section{Introduction} \label{sec:intro}

One of the long lasting dreams in software engineering research is the automated generation of source code. Towards this goal, several approaches have been proposed. The first attempts targeted the relatively simpler problem of code completion, that has been tackled exploiting historical information \cite{Robb2010a}, coding patterns mined from software repositories \cite{Hindle:icse2012,Nguyen:icse2012,Tu:fse2014,Asaduzzaman2014,Nguyen:msr2016,niu2017api,Hellendoorn:fse2017} and, more recently, Deep Learning (DL) models \cite{White2015,Karampatsis:DLareBest,kim2020code,alon2019structural,svyatkovskiy2020intellicode,CiniselliTse2021}.

The release of \Gcopilot \cite{chen2021evaluating} pushed the capabilities of these tools to whole new levels. The large-scale training performed on the OpenAI's Codex model allows Copilot to not limit its recommendations to few code tokens/statements the developer is likely to write: Copilot is able to automatically synthesize entire functions just starting from their signature and natural language descriptions.

This new generation of code recommender systems has the potential to change the way in which developers write code \cite{Ernst:sw2022} and comes with a number of questions concerning how to effectively exploit them to maximize developers' productivity. 

Intuitively, the ability of the developer to provide ``proper'' inputs to the model will become central to boost the effectiveness of its recommendations. In the concrete example of GitHub Copilot, the natural language description provided to the model to automatically generate a code function could substantially influence the model output. This means that two developers providing different natural language descriptions for the same function they would like to automatically generate could receive two different recommendations. While this would be fine in case the two descriptions are actually different in the semantics of what they describe, receiving different recommendations for \emph{semantically equivalent natural language descriptions} would pose questions on the robustness and usability of DL-based code recommenders. 

This is the main research question we investigate in this paper: We study the extent to which different semantically equivalent natural language descriptions of a function result in different recommendations (\ie different synthesized functions) by GitHub Copilot. The latter is selected as representative of DL-based code recommenders since it is the \emph{de facto} state-of-the-art tool when it comes to code generation.

We collected from an initial set of 1,401 open source projects a set of \methods Java methods that are (i) accompanied by a Doc Comment for the Javadoc tool, and (ii) exercised by a test suite written by the project's contributors. Then, as done in the literature \cite{Hu:icpc2018,Li:fse2020}, we considered the first sentence of the Doc Comments as a ``natural language description'' of the method. We refer to this sentence as the ``\emph{original}'' description.

We preliminarily checked whether existing automated paraphrasing techniques are suitable for robustness testing, \ie if they can be used to create semantically equivalent descriptions of the methods to generate. We validated two state-of-the-art approaches in this scenario: PEGASUS \cite{zhang2019pegasus}, a DL-based paraphrasing tool, and Translation Pivoting (TP), a heuristic-based approach. We used both techniques to generate a paraphrase for each \emph{original} description in our dataset. Then, we manually inspected the obtained paraphrases and classified them as semantically equivalent or not. We obtained positive results for both the approaches, with TP being the best performing one with 77\% of valid paraphrases.

Then, to answer our main research question, we generated different paraphrases for each \emph{original} description.

\eject

We used the two previously described automated approaches, \ie PEGASUS and TP, and we also manually generated paraphrases by distributing the original descriptions among four of the authors, each of which was in charge of paraphrasing a subset of them. 

Therefore, for each \emph{original} description, we obtained a set of semantically equivalent \emph{paraphrased} descriptions. We provided both the \emph{original} and the \emph{paraphrased} descriptions as input to \copilot, asking it to generate the corresponding method body. We analyze the percentage of cases in which the \emph{paraphrased} descriptions result in a different code prediction as compared to the \emph{original} one, with a particular focus on the impact on the prediction quality, \eg cases in which the \emph{original} description resulted in the recommendation of a method passing its associated test cases while switching to a \emph{paraphrased} description made \copilot recommending a method failing its related tests.

Our results show that paraphrasing a description results in a change in the code recommendation in $\sim$46\% of cases. The resulting changes also cause substantial variations in the percentage of correct predictions. Such findings indicate the central role played by the model's input in the code recommendation and the need for testing and improving the robustness of DL-based code generators.



Data and code used in our study are publicly available \cite{replication}.

\section{Study Design} \label{sec:study}
The \textit{goal} of our study is to understand how robust is a state-of-the-art DL-based code completion approach (\ie \Gcopilot). We aim at answering the following research questions: \smallskip

\textbf{RQ$_0$: To what extent can automated paraphrasing techniques be used to test the robustness of DL-based code generators?} Not always natural language processing techniques can be used out of the box on software-related text \cite{Lin:icse2018}. Therefore, with this preliminary RQ, we want to understand whether existing automated techniques for generating natural language paraphrases are suitable for SE task at hand (\ie paraphrasing a function description). \smallskip

\textbf{RQ$_1$: To what extent is the output of GitHub Copilot influenced by the code description provided as input by the developer?} This RQ aims at understanding whether \copilot, as a representative of DL-based code generators, is likely to generate different recommendations for different semantically equivalent natural language descriptions provided as input. \smallskip

In the following we detail the context for our study (\secref{sec:context_selection}) and how we collected (\secref{sub:collection}) and analyzed (\secref{sub:analysis}) the data needed to answer our RQs. 

\subsection{Context Selection}
\label{sec:context_selection}
The context of our study is represented by \methods Java methods collected through the following process. We selected all GitHub Java repositories having at least 300 commits, 50 contributors, and 25 stars. These filters have been used in an attempt to exclude personal/toy projects. 

We also excluded forked projects to avoid duplicates. The decision to focus on a single programming language aimed instead at simplifying the non-trivial toolchain needed to run our study. The whole repositories selection process has been performed using the GitHub search tool by Dabic \etal \cite{Dabic:msr2021data}. At this stage, we obtained 1,401 repositories.

In our experimental design, we use the passing/failing tests as a proxy to assess the correctness of the predictions generated by \copilot.
Thus, we need the projects to use a testing framework and to be compilable. We selected all projects that used Maven as build automation tool and for which the build of their latest release succeeded. We obtained 214 repository. By parsing the POM (Project Object Model) file\footnote{POM files are used in Maven to declare dependencies towards libraries.} we only considered projects having as dependencies both jUnit \cite{junit} --- a well-known unit testing framework --- and Jacoco \cite{jacoco} --- a code coverage library. We analyzed the Jacoco reports and selected as methods subject of our experiment those having at least 75\% of statement coverage. 
This gives us confidence that the related test cases exercise an acceptable number of behaviors and, therefore, could allow to spot cases in which different generated functions for semantically-equivalent descriptions actually behave differently. We are aware that passing tests does not imply correctness. We discuss this aspect in \secref{sec:threats}. 

Given our goal to use the method's description as input for \copilot, we also exclude methods not having any associated Doc Comment for the Javadoc tool. Then, we process the Doc Comment of each method in our dataset to extract from it the first sentence (\ie from the beginning to the first ``.''). This is the same approach used in the literature when building datasets aimed at training DL-based techniques for Java code summarization (see \eg \cite{Hu:icpc2018,Li:fse2020}), with the training set composed by pairs $<$\smalltexttt{method, code\_description}$>$, with the latter being the first sentence of the Doc Comment. To ensure that the extracted sentence contains enough wording for the code description, we exclude all methods having less than 10 tokens in the extracted first sentence, since their description may not be sufficient for synthesizing the method.

\begin{table}[h]
		\centering
        \caption{Our dataset of \methods methods from 33 repositories}
	        \begin{tabular}{l|rrr}
	                \toprule
	                & \textbf{Avg} & \textbf{Median} & \textbf{St. Dev.}\\
	                
	                \midrule
					\textbf{\# Tokens}     & 154.3 &  92.0 & 218.2 \\
					\textbf{\# Parameters} &   1.6 &   1.0 &   1.2 \\
					\textbf{\# Cyclomatic Complexity} &   5.3 &   3.0 &   7.6 \\\midrule
	                \textbf{\% Coverage}   &  96.1 & 100.0 &   6.7 \\
	               \bottomrule
	        \end{tabular}
        
    \label{tab:dataset}
\end{table}

\begin{figure}[t]
	\centering
	\includegraphics[width=0.9\linewidth]{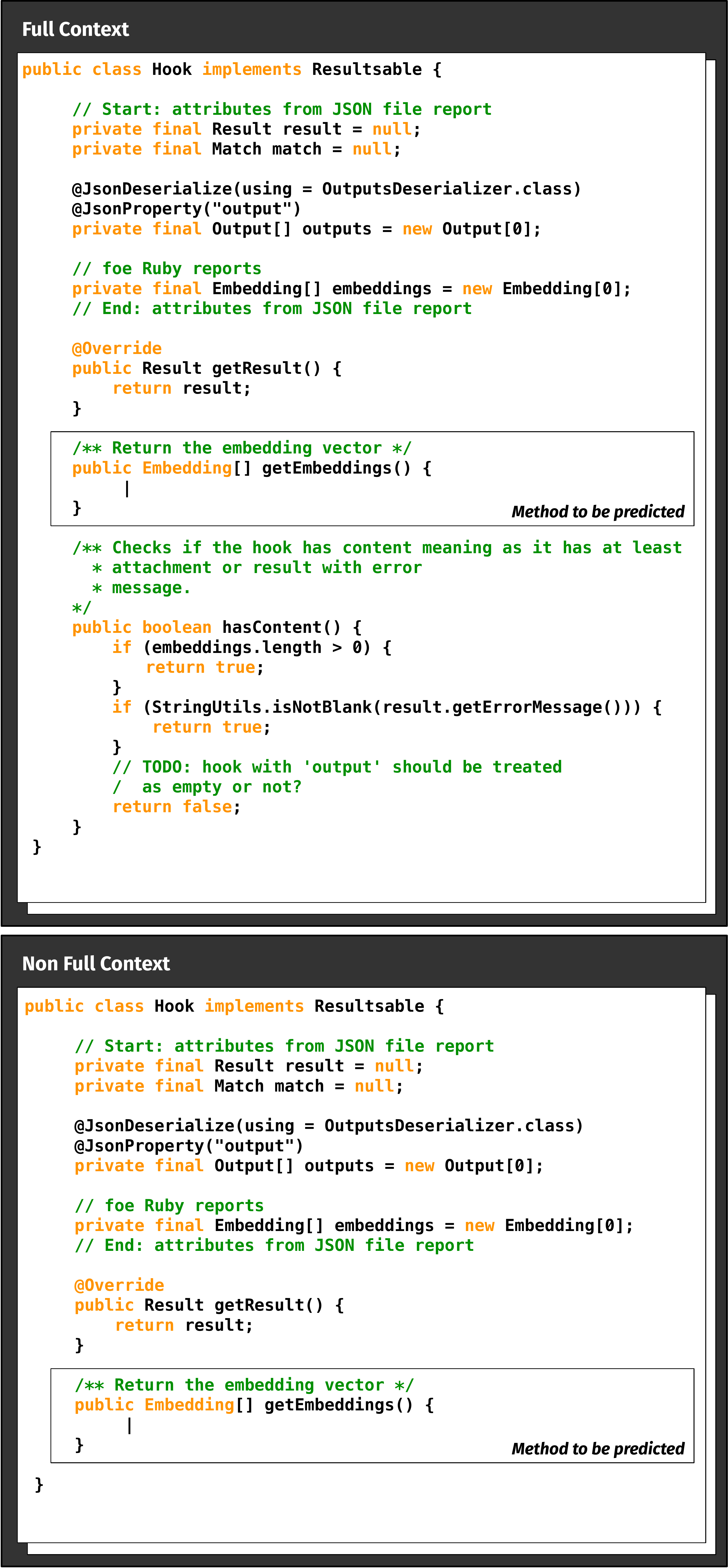}
	\caption{\emph{GitHub Copilot's} input for both code context representations}
	\vspace{-0.3cm}
	\label{fig:context-copilot}
\end{figure}

The above-described process resulted in the collection of \methods Java methods. \tabref{tab:dataset} shows descriptive statistics about their characteristics in terms of number of tokens, parameters and cyclomatic complexity. These three together provide an idea about the complexity of the task \copilot was asked to perform (\ie the complexity of the methods it had to generate). Statistics about the coverage show, instead, the by-design high statement coverage we ensure for the included methods.

\subsection{Data Collection}
\label{sub:collection}

To address \textbf{RQ$_0$}, we experiment with two state-of-the-art paraphrasing techniques. The first is named PEGASUS \cite{zhang2019pegasus}, and it is a sequence-to-sequence DL model pre-trained using self-supervised objectives specifically tailored for abstractive text summarization and fine-tuned for the task of paraphrasing \cite{pegasus}. As for the second technique, we opted for Translation Pivoting (TP). 

Such a technique relies on natural language translation services to translates the \emph{original} description $o$ from English into a foreign language (\ie French), obtaining $o{E \rightarrow F}$. Then, $o{E \rightarrow F}$ is translated back in the original language ($o_{E \rightarrow F \rightarrow E}$) obtaining a paraphrase. 

We provide each technique with the \emph{original} description as input. TP failed to generate a valid paraphrase (\ie a sentence different from the original one) in 100 cases (out of \methods), while this only happened once with PEGASUS. We manually analyzed whether the valid paraphrases we obtained were actually semantically equivalent to the \emph{original} description. For such a process, each of the 1,683 paraphrases (\methods for each of the two tools minus the 101 invalid ones) has been independently inspected by two authors who classified it as semantically equivalent or not. Conflicts, that arisen in 11.9\% (PEGASUS) and 16.54\% (TP) of cases, have been solved by a third author not involved in the first place. 

Concerning \textbf{RQ$_1$}, we start from the \textit{original} description and we generate semantically equivalent descriptions by (i) using the two automated tools, \ie PEGASUS \cite{pegasus} and TP, and (ii)  manually generating paraphrases. For the manual paraphrasing, we split the \methods methods together with their \emph{original} description  into four sets and assigned each of them to one author. Each author was in charge of writing a semantically equivalent but different description of the method by looking at its code and \emph{original} description. This resulted in a dataset (available in \cite{replication}) in which, for each subject method, we have its \emph{original} and \emph{paraphrased} description. 
In the end, for each \emph{original} sentence, we had between one and three paraphrases: \emph{paraphrased}$_{\mathit{PEGASUS}}$, \emph{paraphrased}$_{\mathit{TP}}$, and \emph{paraphrased}$_{\mathit{manual}}$. While \emph{paraphrased}$_{\mathit{manual}}$ is available for all the methods, \emph{paraphrased}$_{\mathit{PEGASUS}}$ and \emph{paraphrased}$_{\mathit{TP}}$ are not. Indeed, we exclude the cases in which each of such tools failed to generate paraphrases (1 and 100, respectively) and the ones that were not considered as semantically equivalent in our manual check (based on the results of RQ$_0$). The maximum number of semantically equivalent paraphrases is 2,575 (up to 891 with PEGASUS, up to 792 with TP, and 892 manually).

The paraphrases, as well as the \textit{original} description, have been used as input to \copilot, simulating developers asking it to synthesize the same Java method by using different natural language descriptions. At the time of our study, \copilot does not provide open APIs to access its services. The only way to use it is through a plugin for one of the supported IDEs. Manually invoking \copilot for the thousands of times needed (up to 6,934, as we will explain later) was clearly not an option. For this reason, we developed a toolchain able to automatically invoke \copilot on the subject instances: We exploit the AppleScript language to automate this task on a MacBook Pro, simulating the developer's interaction with Visual Studio Code (\emph{vscode}). 

For each method $m_i$ in our dataset, we created up to four different versions of the Java file containing it (one for each of the experimented descriptions). In all such versions, we (i) emptied $m_i$'s body, just leaving the opening and closing curly bracket delimiting it; and (ii) removed the Doc Comment, replacing it with one of the four code descriptions we prepared. 

Starting from these files, the automation script we implemented (available in our replication package \cite{replication}) performs the following steps on each file $F_i$.

First, it opens $F_i$ in \emph{vscode} and moves the cursor within the curly brackets of the method $m_i$ of interest. Then, it presses ``\smalltexttt{return}'' to invoke \copilot, waiting up to 20 seconds for its recommendation. Finally, it stores the received recommendation, that could possibly be empty (\ie no recommendation received). To better understand this process, the top part of \figref{fig:context-copilot} depicts how the invocation of \copilot is performed. The gray box represents the whole Java file (\ie the context used by \copilot for the prediction). The emptied method (\ie \smalltexttt{getEmbeddings}) is framed with a black border, with the cursor indicating the position in which \copilot is invoked. The green comment on top of the method represents one of the descriptions we created. As it can be seen, \figref{fig:context-copilot} includes for the same Java file two different scenarios, named \FC and \NFC. In the \FC scenario (top part of \figref{fig:context-copilot}) we provide \copilot with the code \textbf{preceding and following} the emptied method, simulating a developer adding a new method in an already existing Java file. In the \NFC scenario, instead, we only provide as context the code preceding the emptied method (bottom part of \figref{fig:context-copilot}), simulating a developer writing a Java file sequentially and implementing a new method. 

The basic idea behind these two scenarios is that the contextual information provided to \copilot can play a role in its ability to predict the emptied method. Overall, the maximum number of \copilot invocations needed for our study is 6,934 (\methods \textit{original} descriptions plus up to 2,575 paraphrases, each of which for 2 context scenarios). After having collected \copilot's recommendations, we found out that sometimes they did not only include the method we asked to generate, but also additional code (\eg other methods). To simplify the data analysis and to make sure we only consider one recommended method, we wrote a simple parsing tool to only extract from the generated recommendation the first valid method (if any).

\subsection{Data Analysis}
\label{sub:analysis}

Concerning \textbf{RQ$_0$}, we report the number and the percentage of \methods methods for which automatically generated paraphrases (\ie those generated by PEGASUS and by TP) have been classified as semantically equivalent to the \emph{original} description. This provides an idea of how reliable these tools are when used for testing the robustness of DL-based code generators. Also, this analysis allows to exclude from RQ$_1$ automatically generated paraphrases that are not semantically equivalent.

To answer \textbf{RQ$_1$}, we preliminarily assess how far the paraphrased descriptions are from the original ones (\ie the percentage of changed words) by computing the normalized token-level Levenshtein distance \cite{levenshtein1966binary} (NTLev) between the \emph{original} ($d_o$) and any \emph{paraphrased} description ($d_p$):
$$
NTLev(d_o, d_p) = \frac{\mathit{TLev}(d_o, d_p)}{\max({\{}|d_o|, |d_p|\})}
$$

\noindent with $\mathit{TLev}$ representing the token-level Levenshtein distance between the two descriptions. 

While the original Levenshtein distance works at character-level, it can be easily generalized at token-level (each unique token is represented as a specific character). In this case, a token is a word in the text. The normalized token-level Levenshtein distance provides an indication of the percentage of words that must be changed in the \emph{original} description to obtain a \emph{paraphrased} one. 

Then, we analyze the percentage of methods for which the \emph{paraphrased} descriptions result in a different method prediction as compared to the \emph{original} one. When they are different, we also assess how far the methods obtained by using a given \emph{paraphrased} description is from the method recommended when providing the \emph{original} description as input. Also in this case we use the token-level Levenshtein distance as metric. The latter is computed with the same formula previously reported for the natural text descriptions; in this case, however, the tokens are not the words but the Java syntactic tokens. Thus, NTLev indicates in this case the percentage of code tokens that must be changed to convert the method obtained through the \emph{original} description into the one recommended with one of the paraphrases.

Finally, we study the ``quality'' of the recommendations obtained using the different descriptions both in the \FC and \NFC scenarios. Given the sets of methods generated from the \textit{original} description and each of the paraphrasing approach considered, we present the percentages of methods for which \copilot: (i) synthesized a method passing all the related test cases (\textit{PASS}); (ii) synthesized a method that does not pass at least one of the test cases (\textit{FAIL}); (iii) generated an invalid method (\ie with syntactic errors) (\textit{ERROR}); (iv) did not generate any method (\textit{EMPTY}). Syntactic errors have been identified as recommendations for which \emph{Java Parser} \cite{java_parser} did not manage to identify a valid recommended method (\ie cases in which \emph{Java Parser} fails to identify a method node in the AST generated for the obtained recommendation). On top of the passing/failing methods, we also compute the  token-level Levenshtein distance and the CodeBLEU \cite{Ren:codebleu} between each synthesized method and the target one (\ie the one originally implemented by the developers). CodeBLEU measures how similar two methods are. Differently from the BLEU score \cite{papineni2002bleu}, CodeBLEU evaluates the predicted code considering not only the overlapping $n$-grams but also syntactic and semantic match of the two pieces of code (predicted and reference) \cite{Ren:codebleu}.

\subsection{Replication Package}
\label{sub:replication}
The code and data used in our study are publicly available \cite{replication}. In particular, we provide (i) the dataset of manually defined and automatically generated paraphrases; (ii) the AppleScript code used to automate the \copilot triggering; (iii) the code used to compute the CodeBLEU and the Levenshtein distance; (iv) the dataset of \methods methods and related tests used in our study; (v) the scripts used to automatically generate the paraphrased descriptions using PEGASUS and TP; and (vi) all raw data output of our experiments.

\begin{figure*}[!htp]
	\centering
	\includegraphics[width=0.9\linewidth]{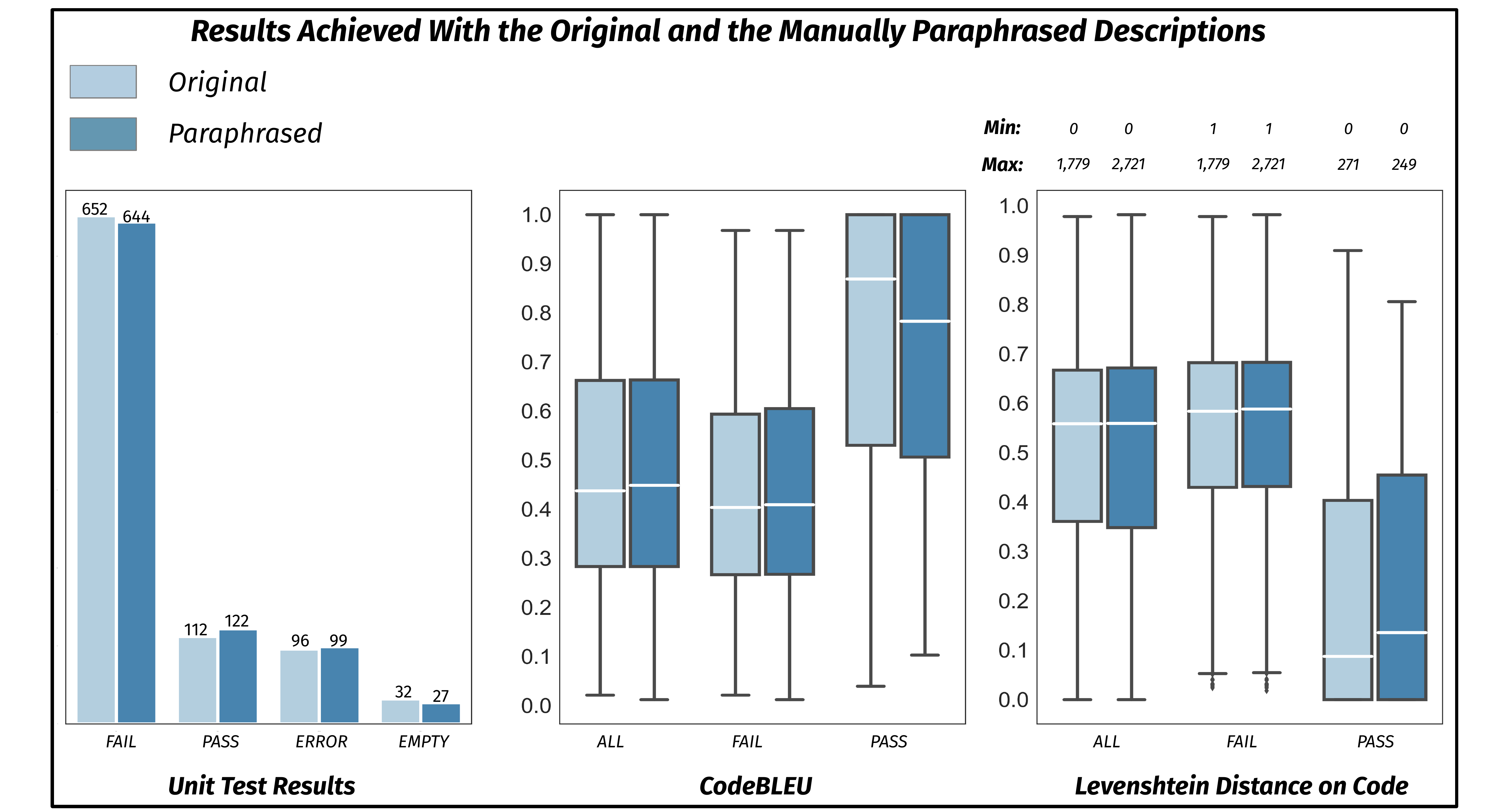}
	\caption{Results achieved by Copilot when considering the \FC code representation on \emph{paraphrases}$_{\mathit{manual}}$.}
	\vspace{-0.2cm}
	\label{fig:full-context-results-developer}
\end{figure*}

\section{Results Discussion} \label{sec:results}

As previously explained, in RQ$_1$ we conducted our experiments both in the \FC and in the \NFC scenario. Since the obtained findings are similar, due to space limitations we only discuss in the paper the results achieved in the \FC scenario (\ie the case in which we provide \copilot with all code preceding and following the method object of the prediction). The results achieved in the \NFC scenario are available in our replication package \cite{replication}.

\subsection{RQ$_0$: Evaluation of Automated Praphrase Generators}

\begin{table}[h]
		\caption{Number of semantically equivalent or nonequivalent paraphrased descriptions obtained using PEGASUS and TP.}
	\resizebox{\linewidth}{!}
	{
	\begin{tabular}{l r r r}
		\toprule
		& Equivalent             & Nonequivalent     & Invalid \\
		\midrule
		PEGASUS & 666 (74.7\%)   & 225 (25.2\%)      & 1 (0.1\%)        \\
		TP      & 688 (77.1\%)   & 104 (11.7\%)      & 100 (11.2\%)     \\
		\bottomrule
	\end{tabular}
	}
	\label{tab:resultsRq2}
\end{table}

\tabref{tab:resultsRq2} reports the number of semantically equivalent and nonequivalent descriptions obtained using the two state-of-the-art paraphrasing techniques, namely PEGASUS and Translation Pivoting (TP), together with the number of invalid paraphrases generated. Out of the 892 \emph{original} descriptions on which they have been run, PEGASUS generated 666 (75\%) semantically equivalent descriptions, while TP went up to 688 (77\%). If we do not consider the invalid paraphrases, \ie the cases for which the techniques do not actually provide any paraphrase, the latter obtains $\sim$87\% of correctly generated paraphrases.

\eject

These findings suggest that the two paraphrasing techniques can be adopted as testing tools to assess the robustness of DL-based code recommenders. In particular, once established a reference description (\eg the \emph{original} description in our study), these tools can be applied to paraphrase it and verify whether, using the reference and the paraphrased descriptions, the code recommenders generate different predictions.

\vspace{0.2cm}
\begin{resultbox}
\textbf{Answer to RQ$_0$.} State-of-the-art paraphrasing techniques can be used as starting point to test the robustness of DL-based code recommenders, since they are able to generate semantically equivalent descriptions of a reference text in up to 77\% of cases.
\end{resultbox}

\subsection{RQ$_1$: Robustness of GitHub Copilot}
\begin{figure*}[!htp]
	\centering
	\includegraphics[width=0.9\linewidth]{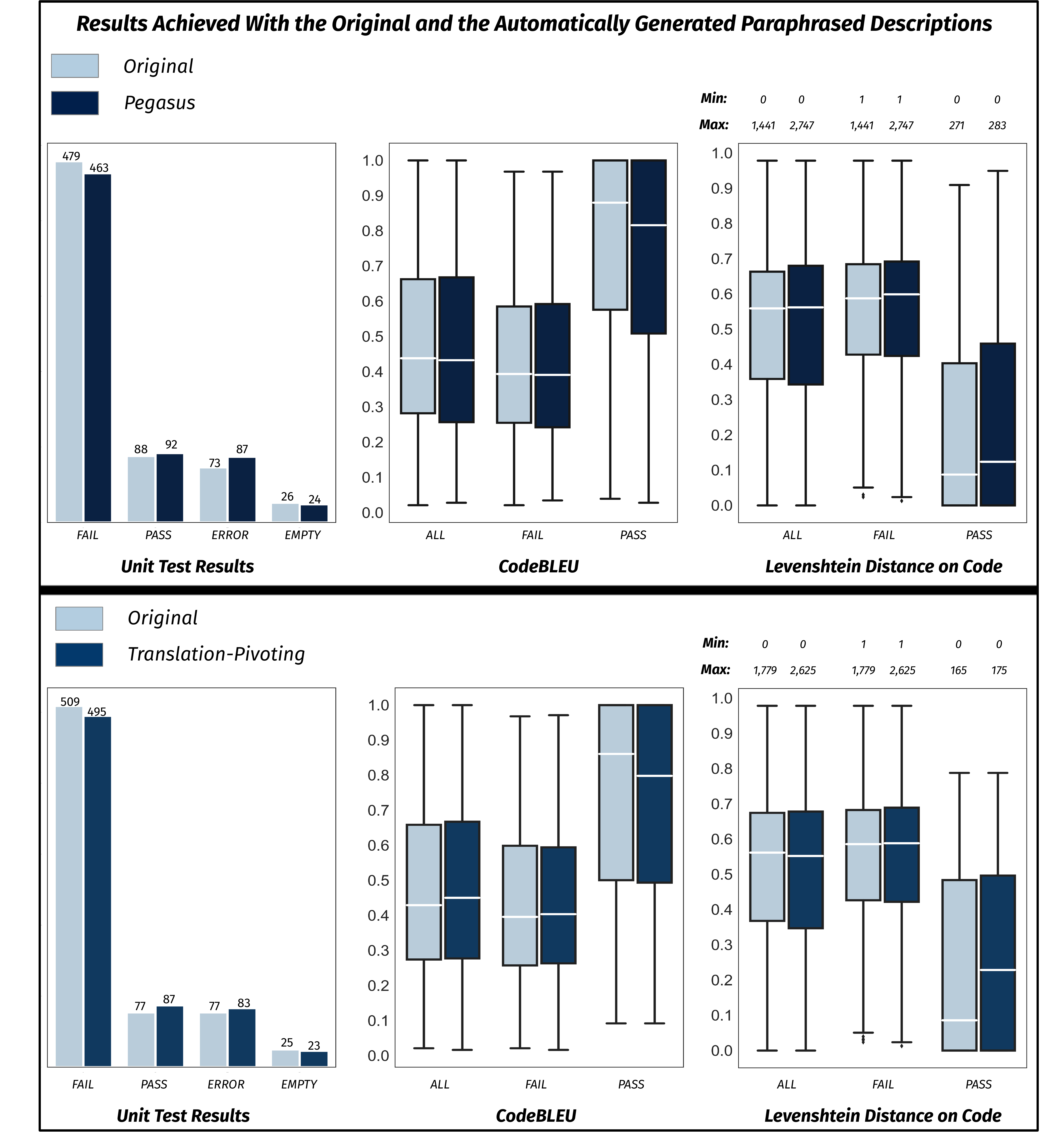}
	\caption{Results achieved by Copilot when considering the \FC code representation on \emph{paraphrases}$_{\mathit{PEGASUS}}$ and \emph{paraphrases}$_{\mathit{TP}}$.}
	\label{fig:full-context-pegasus-pivoting}
\end{figure*}
%
%

\textbf{Performance of Copilot when using the original and the paraphrased description as input.} \figref{fig:full-context-results-developer} summarizes the performance achieved by \copilot when using the \emph{original} description (light blue) and the manually generated \emph{paraphrased} description (dark blue) as input. Similarly, we report in \figref{fig:full-context-pegasus-pivoting} the performance obtained when considering the paraphrases generated with the two automated techniques, \ie PEGASUS and TP (top and bottom of \figref{fig:full-context-pegasus-pivoting}, respectively). It is worth noticing that, in the latter, we only considered in the analysis the paraphrases manually considered as equivalent in RQ$_0$, \ie 666 for PEGASUS and 688 for TP.

A first interesting result is that, as it can be noticed from \figref{fig:full-context-results-developer} and \figref{fig:full-context-pegasus-pivoting}, the results obtained with the three methodologies are very similar. For this reason, to avoid repetitions, in the following, we will mainly focus on the results obtained with the manually generated paraphrases.

Also, as we will discuss, the quality of \copilot's recommendations is very similar when using the \emph{original} and the \emph{paraphrased} descriptions.

In \figref{fig:full-context-results-developer}, the bar chart in the left side reports the number of methods recommended by \copilot (out of 892) that resulted in failing tests, passing tests, syntactic errors, and no (\ie empty) recommendation. Looking at such a chart, the first thing that leaps to the eyes is the high percentage of Java methods ($\sim$73\% for the \emph{original} and $\sim$72\% for the \emph{paraphrased} description) for which \copilot was not able to synthesize a method passing the related unit tests. 

Only $\sim$13\% of instances (112 and 122 depending on the used description) resulted in test-passing methods. While such a result seems to indicate limited performance of \copilot, it must be considered the difficulty of the code generation tasks involved in our study. Indeed, we did not ask \copilot to generate simple methods possibly implementing quite popular routines (\eg a method to generate an MD5 hash from a string) but rather randomly selected methods that, as shown in \tabref{tab:dataset}, are composed, on average, by more than 150 tokens (median = 92) and have an average cyclomatic complexity of 5.3 (median = 3.0). 

Thus, we consider the successful generation of more than 110 of these methods a quite impressive result for a code recommender. The remaining $\sim$15\% of instances resulted either in a parsing error ($\sim$100 methods) or in an empty recommendation ($\sim$30 methods).

The box plot in the middle part of \figref{fig:full-context-results-developer} depicts the results achieved in terms of CodeBLEU \cite{Ren:codebleu} computed between the recommended methods and the target one (\ie the one implemented by the original developers). Higher values indicate higher similarity between the compared methods. Instead, in the right box plot, we show the normalized Levenshtein distance, for which lower values indicate higher similarity. 

For both metrics, we depict the distributions when considering all generated predictions, the ones failing tests, and the ones passing tests. As expected, higher (lower) values of CodeBLEU (Levenshtein distance) are associated with test-passing methods. Indeed, for the latter, the median CodeBLEU is $\sim$0.80 (Levenshtein = $\sim$0.10) as compared to the $\sim$0.40 (Levenshtein = $\sim$0.58) of test-failing methods.  Despite such an expected finding, it is interesting to notice that 25\% of test-passing methods have a rather low CodeBLEU $<$0.50. 

\begin{figure}[tb]
\centering
	\includegraphics[width=0.9\linewidth]{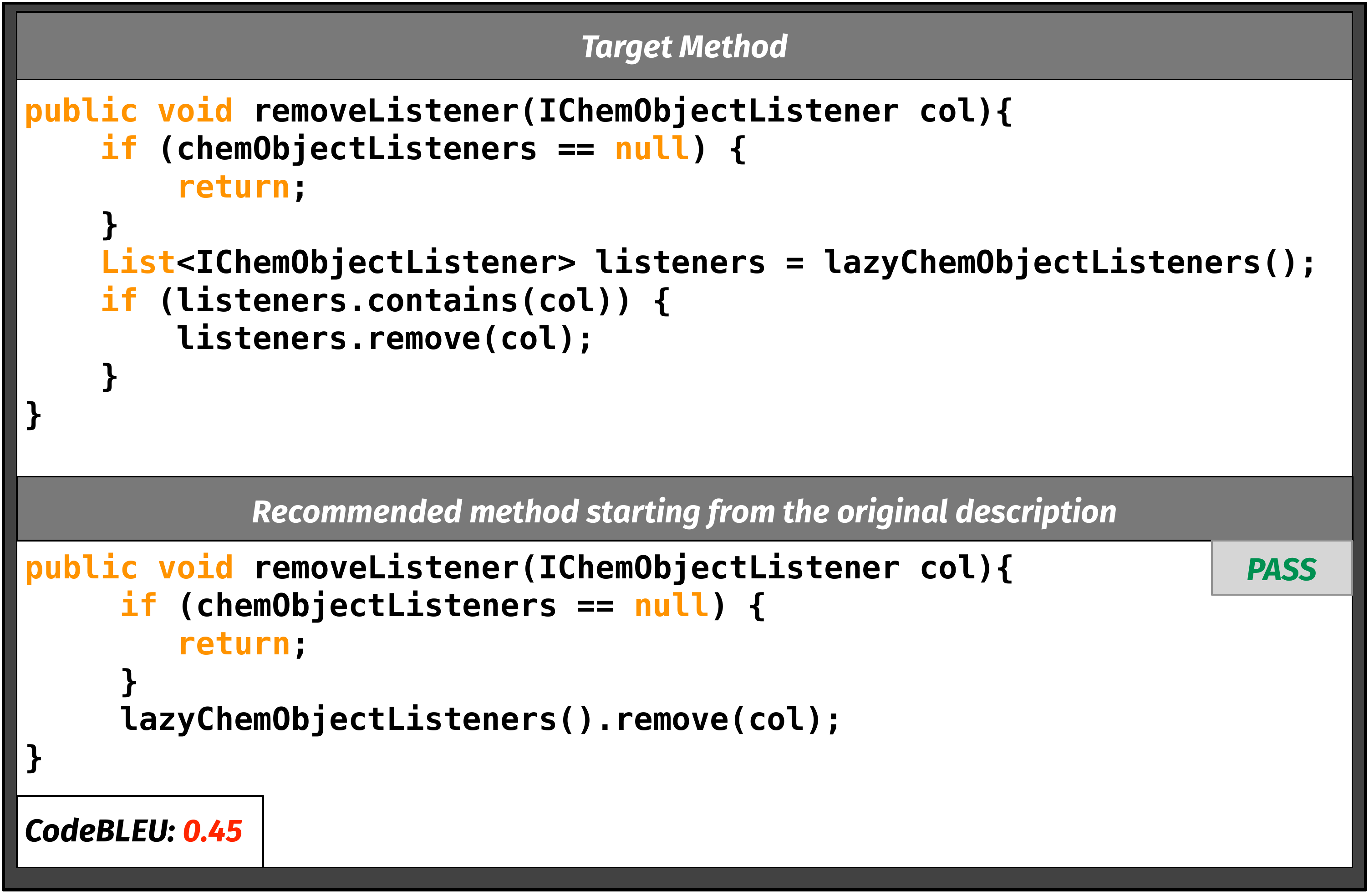}
	\caption{Example of recommended method that passes the unit tests but reports a low CodeBLEU score compared to the oracle (\ie target method).}
	\label{fig:low-code-bleu}
\end{figure}

\figref{fig:low-code-bleu} shows an example of recommended method having a CodeBLEU with the target method of 0.45 and passing the related tests. The recommended method, while substantially different from the target, captures the basic logic implemented in it. The target method first checks if the object \smalltexttt{chemObjectListeners} is \smalltexttt{null} and, if not, it proceeds removing from the \smalltexttt{listeners} list the element matching the one provided as parameter (\ie \smalltexttt{col}). The method synthesized by \copilot avoids the second \smalltexttt{if} statement by directly performing the remove operation after the \smalltexttt{null} check. 

Note that there the two implementations are equivalent: The \smalltexttt{remove} method of \smalltexttt{java.util.List} preliminarily checks whether the passed element is contained in the list before removing it. While the check in the original method has no functional role, together with the introduction of the \smalltexttt{listeners} variable, it might have been introduced to make the method more readable and self-explanatory.

\eject

\begin{figure}[tb]
	\centering
	\includegraphics[width=0.9\linewidth]{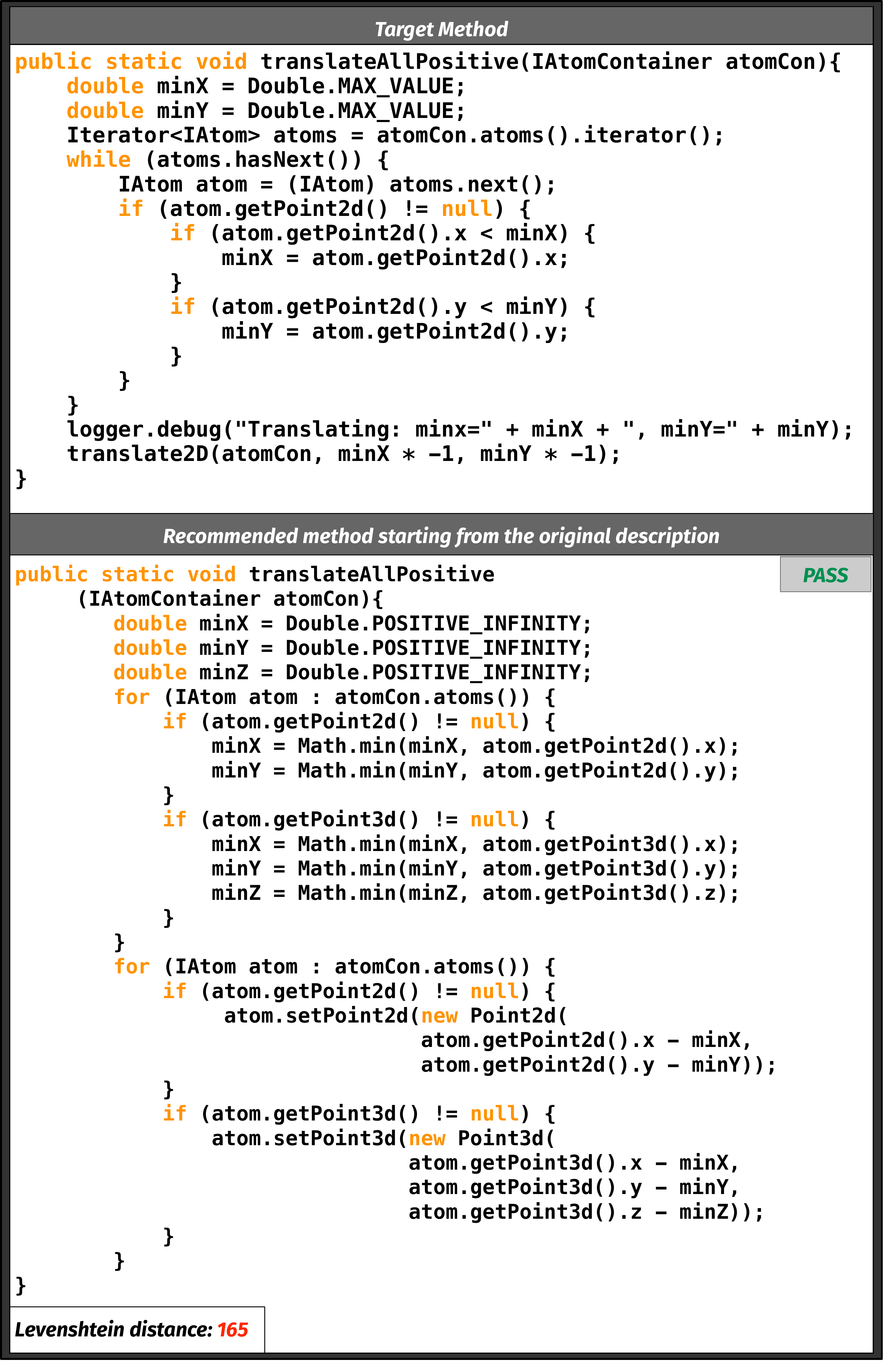}
	\caption{Example of recommended methods that pass the unit tests but would require 165 edit actions to match the target method.}
	\label{fig:high-levenshtein}
\end{figure}


Similarly, \figref{fig:high-levenshtein} shows an example of prediction passing the tests but that, accordingly to the Levenshtein distance, would require 165 token-level edits to match the target prediction (NTLev=63\%). Differently from the previous example, it is clear that, in this case, the two methods do not have the same behavior since the recommended one also treats 3D points, while the original one only 2D points. In other words, the tests fail to capture the difference in the behavior.

These examples provide two interesting observations. The first is that, metrics such as CodeBLEU and Levenshtein distance may result in substantially wrong assessments of the quality of a prediction. Indeed, while the discussed predictions have low CodeBLEU/high Levenshtein values and, thus, would be considered as unsuccessful predictions in most of the empirical evaluations, it is clear that they are valuable recommendations for a developer, even when not 100\% correct (see \figref{fig:high-levenshtein}). This poses questions on the usage of these metrics in the evaluation of code recommenders. Second, also the testing-based evaluation shows, as expected, some limitations as in the second example, in which the two methods do not implement the same behavior but both pass the tests. 

\eject

As a final note, it is also interesting to observe as 25\% of test-failing predictions exhibit high values ($>\sim$0.60) of CodeBLEU, indicating a high code similarity that, however, does not reflect in test-passing recommendations.

\begin{figure*}[h!]
	\centering
	\includegraphics[width=\linewidth]{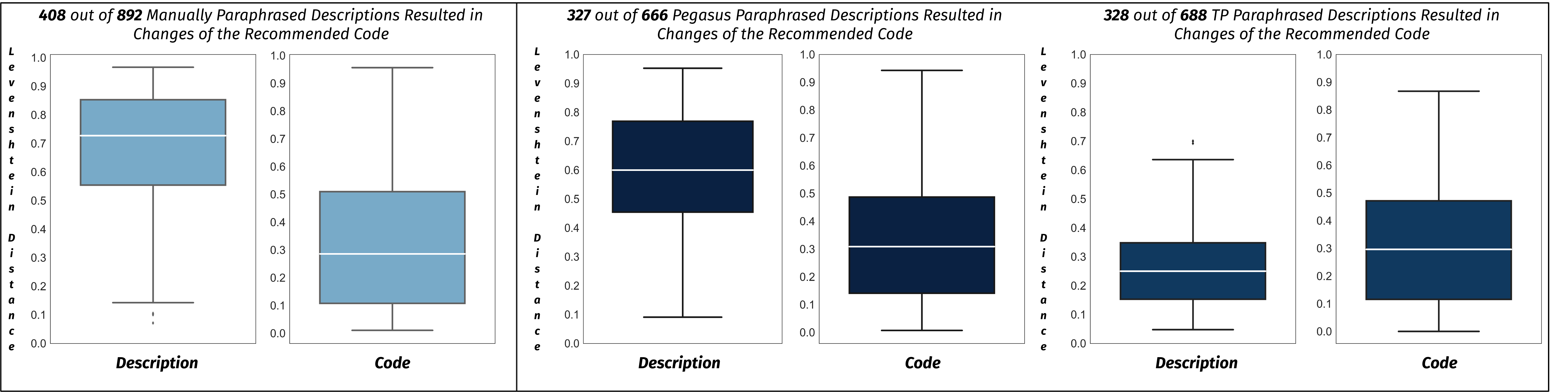}
	\caption{Levenshtein distance between the \emph{original} description and (i) the manually \emph{paraphrased} descriptions (left part) and (ii) the descriptions automatically paraphrased by PEGASUS (middle part) and Translate Pivoting (right). Similarly, we report the Levenshtein distance between the method recommended using the \emph{original} description and the three paraphrases. The latter is only computed for recommendations in which the obtained output differs.}
	\label{fig:changes-overall}
\end{figure*}

\textbf{Impact of paraphrasing the input descriptions.} Out of the 892 manually paraphrased descriptions, 408 (46\%) result in different code recommendations as compared to the \emph{original} description. This means that \copilot synthesizes different methods when it is provided as input with the \emph{original} description and with the manually \emph{paraphrased} description, which are supposed to summarize the same piece of code. Note that at this stage we are not focusing on the ``quality'' of the obtained predictions in any way. We are just observing that different input descriptions have indeed an impact on the recommended code. This implies that developers using different wordings to describe a needed method may end up with different recommendations. Such differences also result in the potential loss of correct recommendations. Indeed, out of the 112 test-passing predictions obtained with the \emph{original} description and the 122 obtained with the manually \emph{paraphrased} description, only 98 are in overlap, indicating that there are 38 correct recommendations only generated either by the \emph{original} (14) or the \emph{paraphrased} (24) description. 

To have a deeper look into the 408 different predictions generated by \copilot with the \emph{original} and the \emph{paraphrased} description, the left part of \figref{fig:changes-overall} (light blue) shows the normalized token-level Levenshtein distance between (i) the \emph{original} description and the \emph{paraphrased} description (see the boxplot labeled with ``Description''), and (ii) the method obtained using the \emph{original} description and that recommended using the \emph{paraphrased} description (``Code''). The ``Description'' boxplot depicts the percentage of words that must be changed to convert the \emph{paraphrased} description into the \emph{original} one. As it can be seen, while describing the same method, the \emph{paraphrased} descriptions can be substantially different as compared to the \emph{original} ones, with 50\% of them requiring changes to more than 70\% of their words. Similarly, the different methods recommended in the 408 cases under analysis, can be substantially different, with a median of $\sim$30\% of code tokes that must be changed to convert the recommendation obtained with the \emph{original} description into the one obtained using the \emph{paraphrased} description (see the ``Code'' boxplot).

These findings are confirmed for the automatically paraphrased descriptions (see the middle and the right  part of \figref{fig:changes-overall} for the results achieved with the PEGASUS and TP paraphrases, respectively). As it can be seen, the main difference as compared to the results of the manually paraphrased description (left part of \figref{fig:changes-overall}) is that TP changes a substantially lower number of words in the \emph{original} description as compared to PEGASUS and to the manual paraphrasing. Such a finding is expected considering that TP just translates the \emph{original} description back and forth from English to French, thus rarely adding new words to the sentence, something that is likely to happen using PEGASUS or by paraphrasing the sentence manually.

\vspace{0.2cm}
\begin{resultbox}
\textbf{Answer to RQ$_1$.} Different (but semantically equivalent) natural language descriptions of the same method are likely to result in different code recommendations generated by DL-based code generation models. Such differences can result in a loss of correct recommendations ($\sim$28\% of test-passing methods can only be obtained either with the \emph{original} or the \emph{paraphrased} descriptions). These findings suggest that testing the robustness of DL-based code recommenders may play an important role in ensuring their usability and in defining possible guidelines for the developers using them. 
\end{resultbox}


\section{Threats to Validity} \label{sec:threats}

Threats to \emph{construct validity} concern the relationship between the theory and what we observe. 
Concerning the performed measurements, we exploit the passing tests as a proxy for the correctness of the recommendations generated by \copilot. We acknowledge that passing tests does not imply code correctness. However, this it can provide hints about the code behavior. To partially address this threat we focused our study on methods having high statement coverage (median = 100\%). Also, we complemented this analysis with the CodeBLEU and the normalized token-level Levenshtein distance. 
As for the execution of our study, we automatically invoked \copilot rather than using it as actual developers would do: We automatically accepted the whole recommendations and did not simulate a scenario in which a developer selects only parts of the provided recommendations. In other words, while our automated script simulates a developer invoking \copilot for help, it cannot simulate the different usages a developer can make of the received code recommendation.

Threats to \emph{internal validity} concern factors, internal to our study, that could affect our results. 
While in RQ$_2$ we had multiple authors inspecting the semantic equivalence of the paraphrasing generated by the automated tools, in RQ$_1$ we relied on a single author to paraphrase the \emph{original} description. This introduces some form of subjectivity bias. However, the whole point of our paper is that, indeed, subjectivity plays a role in the natural language description of a function to generate and we are confident that the written descriptions were indeed semantically equivalent to the \emph{original} one. Indeed, the authors involved in the manual paraphrasing have an average of seven years of experience in Java.
Also related to internal validity is our choice of using the first sentence of the Doc Comments as the \emph{original} natural language description. These sentences may be of low quality and not representative of how a developer would describe a method they want to automatically generate. This could substantially influence our findings, especially in terms of the effectiveness of \copilot (\ie its ability to generate test-passing methods). However, such a threat is at least mitigated by the fact that \copilot has also been invoked using the manually written descriptions, showing a similar effectiveness.
A final threat regards the projects used for our study.

Those are open-source projects from GitHub, and it is likely that at least some of them have been used for training Copilot itself. In other words, the absolute actual effectiveness reported might not be reliable. However, the objective of our study is to understand the differences when different paraphrases are used rather than the absolute performance of Copilot, like previous studies did (\eg \cite{nguyen2022empirical}).

Threats to \emph{external validity} are related to the possibility to generalize our results. Our study has been run on \methods methods we carefully selected as explained in \secref{sec:context_selection}. Rather than going large-scale, we preferred to focus on methods having a high test coverage and a verbose first sentence in the Doc Comment. Larger investigations are needed to corroborate or contradict our findings. Similarly, we only focused on Java methods, given the effort required to implement the toolchain needed for our study, and in particular the script to automatically invoke \copilot and parse its output. Running the same experiment with other languages is part of our future agenda.

\section{Related Work} \label{sec:related}

%

Recommender systems for software developers are tools supporting practitioners in daily activities~\cite{McMillan:tse2013, robillard:recommenders}, such as documentation writing and retrieval~\cite{Xie:msr2006,Moreno:icse2015,Moreno:tse2016,Xing:icpc2018}, refactoring~\cite{Bavota:emse2014,Tsantalis:saner2018}, bug triaging~\cite{Tamrawi:2011,Xia:tse2016}, bug fixing~\cite{goues:icse2012,Tufano:tosem2019,Li:icse2020}, \etc Among those, code recommenders, such as code completion tools, have became a crucial feature of modern Integrated Development Environments (IDEs) and support in speeding up code development by suggesting the developers code they are likely to write~\cite{Bruch:fse2009,kim2020code,Ciniselli2021}. Given the empirical nature of our work, that focuses on investigating a specific aspect of code recommenders, in this section we do not discuss all pervious works proposing novel or improving existing code recommenders (see \eg~\cite{Xie:msr2006, Moreno:icse2015, Moreno:tse2016, goues:icse2012, Tufano:tosem2019, Li:icse2020, Wen:icse2021, Nguyen:fse2016, Liu:ase2020, Kim:ASE2009, Allamanis:fse2014, kim2020code, Karampatsis:DLareBest, Watson:icse2020, Tufano:asserts}). Instead, we focus on empirical studies looking at code recommenders from different perspectives (\secref{sub:rel1}) and on studies specifically focused on GitHub Copilot (\secref{sub:rel2}).

\subsection{Empirical Studies on Code Recommenders}
\label{sub:rel1}
Proksch \etal~\cite{proksch2016evaluating} conducted an empirical study aimed at evaluating the performance of code recommenders when suggesting method calls. Their study has been run on a real-world dataset composed of developers' interactions captured in the IDE. Results showed that commonly used evaluation techniques based on synthetic datasets extracted by mining released code underperform due to a context miss. 

On a related research thread, Hellendoorn \etal~\cite{hellendoorn2019code} compared code completion models on both real-world and synthetic datasets.
Confirming what observed by Proksch \etal, they found that the evaluated tools are less accurate on the real-world dataset, thus concluding that synthetic benchmarks are not representative enough. Moreover, they found that the accuracy of code completion tools substantially drops in challenging completion scenarios, in which developers would need them the most.

M{\u{a}}r{\u{a}}șoiu \etal~\cite{muaruasoiu2015empirical} analyzed how practitioners rely on code completion during software development.
The results showed that the users actually ignore many synthesized suggestions. Such a finding has been corroborated by Arrebola and Junior~\cite{arrebola2017source}, who stressed the need for augmenting code recommender systems with the development's context.

Jin and Servant~\cite{jin2018hidden} and Li \etal \cite{li2021toward} investigated the \textit{hidden costs} of code recommendations. Jin and Servant found that IntelliSense, a code completion tool, sometimes underperforms by providing the suitable recommendation far from the top of the recommended list of solutions. Consequently, developers are discouraged from picking the right suggestion. Li \etal, aware of this potential issue, conducted a coding experiment in which they try to predict whether correct results are generated by code completion models, showing that their approach can reduce the percentage of false positives up to 70\%. 

Previous studies also assessed the actual usefulness of these tools. Xu \etal~\cite{xu2021inide} ran a controlled experiment with 31 developers who were asked to complete implementation tasks with and without the support of two code recommenders. They found a marginal gain in developers' productivity when using the code recommenders.

Ciniselli \etal~\cite{CiniselliTse2021} empirically evaluated the performance of two state-of-the-art Transformer-based models in challenging coding scenarios, for example, when the code recommender is required to generate an entire code block (\eg the body of a \smalltexttt{for} loop). The two experimented models, RoBERTa and Text-To-Text Transfer Transformer (T5), achieved good performance ($\sim$69\% of accuracy) in the more classic code completion scenario (\ie predicting few tokens needed to finalize a statement), while reported a substantial drop of accuracy ($\sim$29\%) when dealing with the previously described more complex block-level completions.

Our study is complementary to the ones discussed above. Indeed, we investigate the robustness of DL-based code recommenders supporting what it is know in the literature as ``\emph{natural language to source code translation}''. We show that semantically equivalent code descriptions can result in different recommendations, thus posing questions on the usability of these tools.  

\subsection{Empirical Studies on GitHub Copilot}
\label{sub:rel2}

GitHub \copilot has been recently introduced as the state-of-the-art code recommender, and advertised as an ``AI pair programmer''~\cite{copilot,howard2021github}. Since its release, researchers started investigating its capabilities. 

Most of the previous research aimed at evaluating the impact of GitHub \copilot on developers' productivity and its effectiveness (in terms of correctness of the provided solutions). 
Imai \cite{imai2022github} investigated to what extent \copilot is actually a valid alternative to a human pair programmer. They observed that \copilot results in increased productivity (\ie number of added lines of code), but decreased quality in the produced code.
Ziegler \etal \cite{ziegler2022productivity} conducted a case study in which they investigated whether usage measurements about \copilot can predict developers' productivity. They found that the acceptance rate of the suggested solutions is the best predictor for perceived productivity.
Vaithilingam \etal \cite{vaithilingam2022expectation} ran an experiment with 24 developers to understand how \copilot can help developers complete programming tasks. Their results show that \copilot does not improve the task completion time and success rate. However, developers report that they prefer to use \copilot because it recommends code that can be used as a starting point and saves the effort of searching online.

Nguyen and Nadi \cite{nguyen2022empirical} used LeetCode questions as input to \copilot to evaluate the solutions provided for several programming languages in terms of correctness --- by running the test cases available in LeetCode --- and understandability --- by computing their Cyclomatic Complexity and Cognitive Complexity \cite{campbell2018cognitive}. They found notable differences among the programming languages in terms of correctness (between 57\%, for Java, and 27\%, for JavaScript). On the other hand, \copilot generates solutions with low complexity for all the programming languages.
While we also measure the effectiveness of the solutions suggested by \copilot, our main focus is on understanding its robustness when different inputs are provided.

Two previous studies aimed at evaluating the security of the solutions recommended by \copilot. 
Hammond \etal \cite{pearce2021empirical} investigated the likelihood of receiving from \copilot recommendations including code affected by security vulnerabilities. They observed that vulnerable code is recommended in 40\% of cases out of the completion scenarios they experimented with.
On a similar note, Sobania \etal~\cite{sobania2021choose} evaluated GitHub \copilot on standard program synthesis benchmark problems and compared the achieved results with those from the genetic programming literature. The authors found that the performance of the two approaches are comparable. However, approaches based on genetic programming are not mature enough to be deployed in practice, especially due to the time they require to synthesize solutions.
In our study, we do not focus on security, but only on the correctness of the suggested solutions.


Albert Ziegler, in a blog post about GitHub \copilot\footnote{\url{https://docs.github.com/en/github/copilot/research-recitation}} investigated the extent to which the tool suggestions are copied from the training set they used. Ziegler reports that \copilot rarely recommends verbatim copies of code taken from the training set.

\section{Conclusions and Future Work} \label{sec:conclusion}

We investigated the extent to which DL-based code recommenders tend to synthesize different code components when starting from different but semantically equivalent natural language descriptions. We selected \Gcopilot as the tool representative of the state-of-the-art and asked it to generate \methods non-trivial Java methods starting from their natural language description. For each method in our dataset we asked \copilot to synthesize it using: (i) the \emph{original} description, extracted as the first sentence in the Javadoc; and (ii) \emph{paraphrased} descriptions. We did this both by manually modifying the \emph{original} description and by using automated paraphrasing tools, after having assessed their reliability in this context.

We found that in $\sim$46\% of cases semantically equivalent but different method descriptions result in different code recommendations. We observed that some correct recommendations can only be obtained using one of the semantically equivalent descriptions as input.


Our results highlight the importance of providing a proper code description when asking DL-based recommenders to synthesize code. In the new era of AI-supported programming, developers must learn how to properly describe the code components they are looking for to maximize the effectiveness of the AI support.

Our future work will focus on answering our first research question \emph{in vivo} rather than \emph{in silico}. In other words, we aim at running a controlled experiment with developers to assess the impact of the different code descriptions they write on the received recommendations. Also, we will investigate how to customize the automatic paraphrasing techniques to further improve their performance on software-related text (such as methods' descriptions).

\section*{Acknowledgments}
This project has received funding from the European Research Council (ERC) under the European Union's Horizon 2020 research and innovation programme (grant agreement No. 851720).

\bibliographystyle{IEEEtranS}
\bibliography{main}

\end{document}